\documentclass[sigconf]{acmart}
\AtBeginDocument{%
  }

\copyrightyear{2026}
\acmYear{2026}
\setcopyright{cc}
\setcctype{by}
\acmConference[SIGIR '26] {Proceedings of the 49th International ACM SIGIR Conference on Research and Development in Information Retrieval}{July 20--24, 2026}{Melbourne, VIC, Australia.}
\acmBooktitle{Proceedings of the 49th International ACM SIGIR Conference on Research and Development in Information Retrieval (SIGIR '26), July 20--24, 2026, Melbourne, VIC, Australia}
\acmISBN{979-8-4007-2599-9/2026/07}
\acmDOI{10.1145/3805712.3808556}

\usepackage{booktabs}
\usepackage{subcaption}
\usepackage{placeins} 
\usepackage{float}
\usepackage{caption} 
\usepackage{hyperref}
\usepackage{diagbox}   % for \diagbox
\usepackage{cuted}   % provides the strip environment
\usepackage{colortbl}
\usepackage{xcolor}
\usepackage{balance}
\definecolor{lightyellow}{RGB}{255, 255, 200}
\newcommand{\deltan}[1]{_{\raisebox{-1.1ex}{\clap{\scriptsize{#1}}}}}

\newcommand{\deltav}[1]{_{\raisebox{-0.8ex}{\clap{\scriptsize{#1}}}}}

\newcommand{\deltas}[1]{%
  _{\raisebox{-0.8ex}{\kern-1.2em\scriptsize #1}}%
}

\newcommand{\randomdoc}{\includegraphics[height=1.8ex]{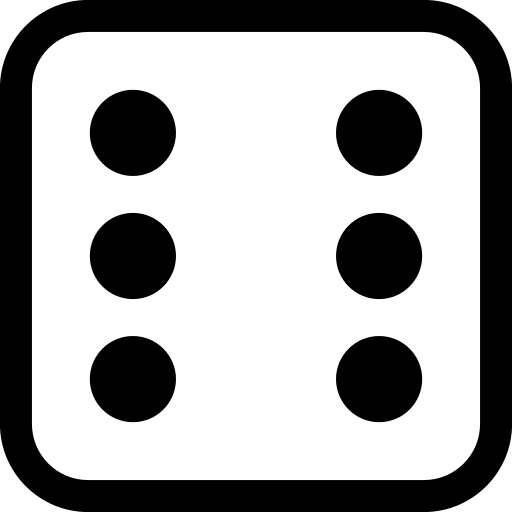}}
\newcommand{\relevantdoc}{\includegraphics[height=1.8ex]{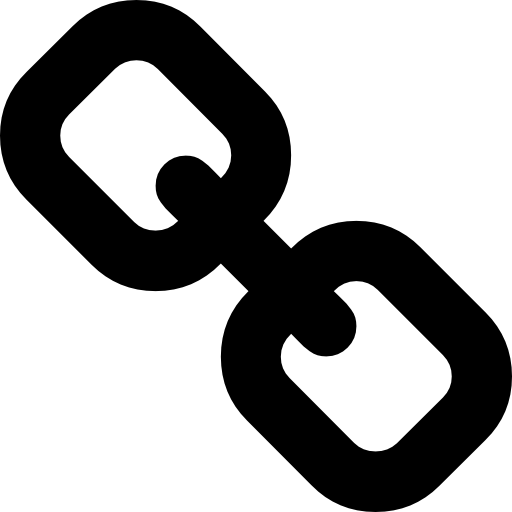}}
\newcommand{\distractdoc}{\includegraphics[height=1.8ex]{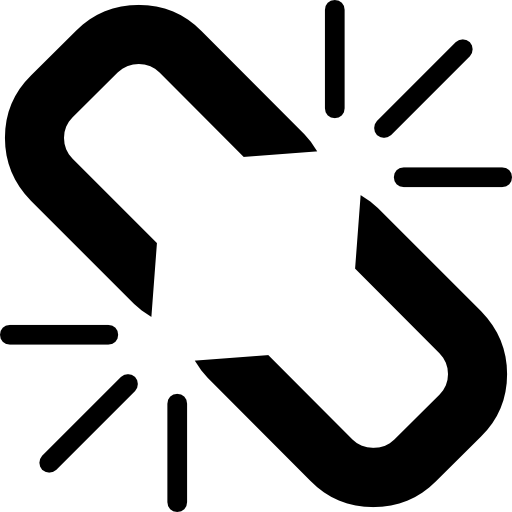}}
\newcommand{\golddoc}{\includegraphics[height=1.8ex]{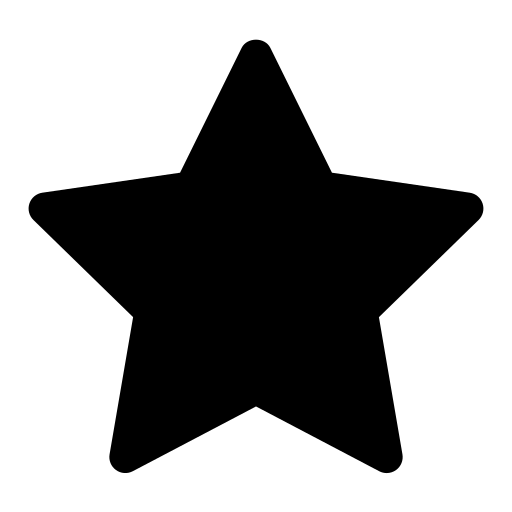}}
\newcommand{\fulldoc}{\includegraphics[height=2ex]{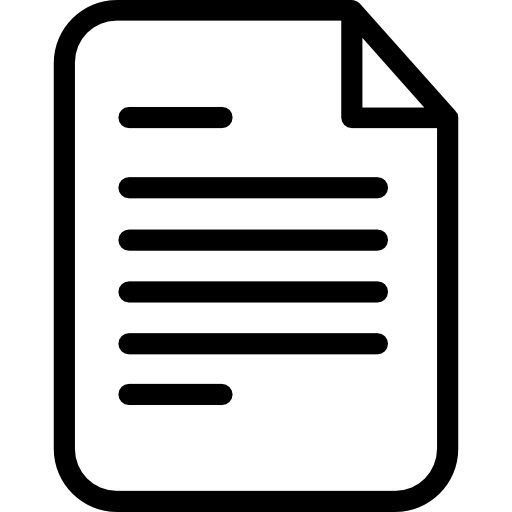}}

\usepackage{footmisc}

\usepackage{enumerate}
\usepackage[shortlabels]{enumitem}

\usepackage{pifont}
\usepackage{makecell} % for stacked cells
\usepackage{amsmath}
\usepackage{tabularx}
\usepackage{caption}
\usepackage{tcolorbox}

\begin{document}

% \title{The Name of the Title Is Hope}
\title{The Powerless Noise: How Experimental Settings Shape the Reported Power of Noise}

\author{Michał Mazuryk}
\affiliation{%
  \institution{University of Amsterdam}
  \city{Amsterdam}
  \country{The Netherlands}
}
\email{michal.mazuryk@student.uva.nl}

\author{Fleur Dolmans}
\affiliation{%
  \institution{University of Amsterdam}
  \city{Amsterdam}
  \country{The Netherlands}
}
\email{fleur.dolmans@student.uva.nl}

\author{Louis Gehringer}
\affiliation{%
  \institution{University of Amsterdam}
  \city{Amsterdam}
  \country{The Netherlands}
}
\email{louis.gehringer@student.uva.nl}

\author{Ina Klaric}
\affiliation{%
  \institution{University of Amsterdam}
  \city{Amsterdam}
  \country{The Netherlands}
}
\email{ina.klaric@student.uva.nl}

\author{Jia-Huei Ju}
\affiliation{%
  \institution{University of Amsterdam}
  \city{Amsterdam}
  \country{The Netherlands}
}
\email{j.ju@uva.nl}

\author{Mohammad Aliannejadi}
\affiliation{%
  \institution{University of Amsterdam}
  \city{Amsterdam}
  \country{The Netherlands}
}
\email{m.aliannejadi@uva.nl}

\renewcommand{\shortauthors}{Michał Mazuryk et al.}

\begin{abstract}
    Recent work has suggested that adding irrelevant documents to the input of retrieval-augmented generation (RAG) systems can improve question-answering performance, a phenomenon referred to as the ``\textit{Power of Noise}.'' This motivated investigations into the role of noise in information retrieval. In this paper, we reproduce the main findings of Cuconasu et al. \cite{cuconasu2024power} and evaluate the robustness of the effect under extended experimental settings. We first confirm that the phenomenon holds under the original setup, which uses earlier-generation LLMs, restrictive prompting and constrained decoding settings. We subsequently introduce a series of extensions to investigate the underlying causes of the noise effect, examining the  authors' original design choices including the use of different models, instruction prompting, and relaxed output length constraints. Across these ablations, the Power-of-Noise pattern proves highly sensitive to inference configuration: it can appear, weaken, or disappear under small changes to prompt formulation and decoding limits. Combined with our error analysis, which shows substantial contributions from truncation and malformed generations, this variance indicates that the original effect cannot be robustly confirmed as a general benefit of noisy retrieval under these experimental conditions. More broadly, our work highlights the importance of carefully scrutinizing inference design in retrieval-augmented generation systems. 
    Our code is available at https://github.com/ina0105/The-Power-of-Noise-Reproduction.
\end{abstract}







\maketitle

\section{Introduction}

Retrieval-Augmented Generation (RAG) \cite{lewis2020retrieval} has become a central method for connecting large language models (LLMs) with current, domain-specific information \cite{chen2024benchmarking, wu2025pandora}. By adding retrieved documents to the model prompts, RAG incorporates external knowledge thereby mitigating common LLM limitations such as hallucinations and outdated information \cite{ 
cao2025evaluating, chen2024benchmarking, lewis2020retrieval, wu2025pandora}. RAG combines retrieval and generation, but its effectiveness depends critically on retrieval quality: irrelevant or incorrect retrieved content can degrade model performance \cite{cao2025evaluating, chen2024benchmarking, cuconasu2024power}. Given that retrieval often struggles in practical settings, developing RAG systems that remain reliable under imperfect retrieval conditions has become crucial.

In \textit{The Power of Noise: Redefining Retrieval for RAG Systems}, Cuconasu et al. \cite{cuconasu2024power} study how distracting, random and realistically retrieved documents influence RAG performance on question answering tasks. While semantically similar but answer-less documents predictably reduce accuracy, they report a surprising result: adding documents that are highly unrelated to the query (i.e., random) can improve LLM performance.  In some settings, this additional context enables models to better identify and extract the correct answer, outperforming inputs that only include the relevant document containing the correct answer. Further experiments show that appending random documents can yield accuracy gains of up to $\sim35\%$, challenging the common assumption that imperfect retrieval necessarily degrades performance \cite{chen2024benchmarking, wang2025astute}. Based on these findings, the authors conclude that 
\textit{random documents seem to exert a positive influence on
LLM accuracy.}
%\textit{a well-designed RAG system should intentionally add random noise to its prompt to improve LLM performance}.
We refer to this phenomenon as the \textit{Power-of-Noise}. 

While the original study presents compelling empirical evidence for the effect, a closer examination reveals several methodological and conceptual limitations that warrant further investigation.

First, the evaluation is restricted to low-parameter LLMs running at 4-bit precision. Strong quantization can significantly distort attention dynamics by compressing key–query dot product values, altering softmax distributions, and introducing rounding artifacts in softmax computation \cite{dettmers2022llmint88bitmatrixmultiplication,frantar2023gptqaccurateposttrainingquantization,xiao2024smoothquantaccurateefficientposttraining}. This is particularly relevant as the authors hypothesize that the noise effect is due to entropy collapse within attention mechanisms~\cite{zhai2023stabilizing}, where numerical precision plays a crucial role. Since entropy collapse is highly sensitive to small perturbations in attention logits, quantization may artificially induce or suppress the reported effect. The study is further limited by the use of earlier-generation LLMs, raising questions about whether the effect holds in modern architectures.

Second, the conclusions rely on a constrained prompting setup. The experiments use a single prompt template in which task instructions and retrieved documents are concatenated as raw text, tokenized and fed to the model without applying any model-specific chat or instruction template, such that chat-tuned models (e.g. Llama-2-chat) are effectively evaluated as plain language models rather than in their intended dialogue format. 

Third, the task instruction itself may introduce artifacts. The prompt requests answers of at most five tokens and provides a \texttt{NO-RES} option when an answer cannot be produced; however, only the generation cap of 15 tokens enforces output length in practice. As a result, the five-token instruction does not strictly constrain responses and may instead trigger undesirable behavior, including hallucinations, as the model attempts to satisfy conflicting instructions. The \texttt{NO-RES} directive may further bias models toward unwarranted refusals even when the answer is present in the context. Prior work has shown that prompt design strongly influences LLM performance and reliability \cite{chen2025unleashing}, and that models tend to forget initial instructions as context length grows \cite{liu-etal-2024-lost}, making these limitations particularly relevant. Combined with exact-match evaluation, these factors can artificially deflate measured accuracy by penalizing truncated, refused, or otherwise correct responses \cite{wang2023evaluating}.

\paragraph{Goal.} This study aims to assess the robustness and generality of the Power-of-Noise effect across model architectures and experimental settings, and to clarify its implications for RAG system design. \\

Given the implications of the original findings for RAG system design, it is important to carefully examine how well these claims hold under broader and more realistic settings. This motivates the following research questions:

\begin{enumerate}[start=1,label={\bfseries Q\arabic*}]
\item How does LLM architecture and numerical precision influence the impact of random noise in RAG? \label{Q1}

% \item What role do decoding setup play in shaping the Power-of-Noise effect? \label{Q2}
\item How does the use of instruction templates affect model robustness under noisy retrieval? \label{Q2}

% \item  How does prompting structure and formation affect LLM behavior under noisy retrieval? \label{Q3}
\item How do generation constraints and prompt design that encourage short answers affect observed performance? \label{Q3}

\end{enumerate}

We reproduce key experiments from the original study using Llama2 \cite{touvron2023llama} and MPT \cite{MosaicML2023MPT7B}, and extend them across model architectures, numerical precision settings, prompting strategies, and decoding configurations. We further evaluate modern instruction-tuned models, including Llama3 \cite{dubey2024llama3}, Mistral \cite{jiang2023mistral7b}, Falcon3 \cite{Falcon3}, and Qwen2.5 \cite{yang2025qwen2}. To better reflect modern RAG practice, we adopt instruction-style prompting, remove restrictive prompt output constraints, increase generation limits to 100 tokens, and control for quantization effects.

Our results show that while the Power-of-Noise effect is reproducible under the original constrained setup, it is not robust: once prompting and decoding configurations are relaxed, baseline performance improves and the apparent benefit of adding random documents weakens or disappears. In particular, the effect vanishes for several modern instruction-tuned models and weakens substantially for Llama3, indicating that the reported gains from random noise largely stem from inference time configurations rather than representing a general property of RAG systems.

% Our contributions are as follows:
% \begin{itemize}
%     \item We reproduce the original experiments under the reported settings to verify the Power-of-Noise effect.
%     \item We assess the robustness and generalizability of the effect on modern instruction-tuned models and under varying numerical precision.
%     \item We conduct systematic ablations on prompt formulation and decoding constraints to identify key experimental factors influencing the observed phenomenon.
% \end{itemize}

\section{Related Work}
Chen et al. \cite{chen2024benchmarking} introduce the \textit{Retrieval-Augmented Generation Benchmark}, showing that state-of-the-art LLMs remain sensitive to irrelevant context, with performance degrading as noise increases. Traditionally viewed as a nuisance to be mitigated, noise has more recently become the focus of work exploring its potential benefits.

Abdolazimi et al. \cite{abdolazimi2024harnessing} survey methods that intentionally introduce noise, highlighting its role in improving robustness, stability, and information extraction. Similarly, Wu et al. \cite{wu2025pandora} challenge the assumption that all noise is harmful in RAG. They propose a linguistically grounded taxonomy of seven noise types and introduce \textit{NoiserBench}, demonstrating that certain noise categories can improve performance across datasets and reasoning tasks. Complementing this line of work, Amiraz et al. \cite{amiraz2025distracting} study how irrelevant passages distract LLMs during generation, formalizing and quantifying distraction effects that are consistent across models.

Beyond noise-centric analyses, prior work has emphasized the sensitivity of LLMs to prompt formulation and interface-level design choices. This sensitivity is particularly pronounced for instruction-tuned models, which rely on explicit task descriptions and structured prompting to correctly interpret retrieval-augmented inputs \cite{sun2023evaluating, sclar2023quantifying}. Even minor variations in prompt wording or task framing can lead to unstable reasoning, reduced consistency, or failure to use retrieved evidence \cite{sclar2023quantifying, zhu2023promptrobust}.

Inference-time constraints further interact with these effects. Restrictive decoding settings, such as aggressive length limits, can truncate valid reasoning or suppress factual accuracy \cite{xia2025controllable}. Similarly, the ordering and formatting of retrieved passages influence how models allocate attention, with earlier work showing uneven use of evidence depending on presentation \cite{liu-etal-2024-lost}. Instruction tuning may amplify the importance of these factors by training models to closely follow natural-language task specifications \cite{zhang2023instruction}. Surveys of prompting strategies, including zero-shot, few-shot, and dynamic prompting, demonstrate substantial performance variation across NLP tasks \cite{sahoo2024systematic}. In practice, RAG systems commonly rely on instruction-style prompts, multi-stage prompting, and chain-of-thought reasoning to cope with imperfect retrieval \cite{wei2024instructrag}.

Despite these works, the role of prompt formulation, instruction structure, and decoding constraints has received relatively little attention in prior research of noise in RAG systems. Our study aims to bridge this gap by examining how such inference-time design choices interact with the reported Power-of-Noise effect.

\section{Methodology}

We first reproduce the experimental setup of Cuconasu et al. \cite{cuconasu2024power} to validate the reported Power-of-Noise under the original conditions. Next, we extend this setup to investigate the robustness of the effect under modern model architectures, prompting strategies, and inference configurations. This two-stage methodology; reproduction followed by controlled extensions, allows us to assess whether the Power-of-Noise reflects a fundamental property of RAG systems or an artifact of experimental design choices.

\subsection{Dataset}
Mimicking the original study, we conducted all experiments on the Natural Questions (NQ) open dataset \cite{kwiatkowski2019natural} which is a subset of the full NQ dataset. NQ-open is large-scale collection of real-world queries derived from Google search data, where the restriction of linking answers to specific Wikipedia passages was removed. Adapting the authors, English Wikipedia dump as of 20 December 2018 is employed as the answering queries, for which each Wikipedia article was segmented into non-overlapping passages of 100 words. To mitigate any potential temporal mismatch between the Wikipedia dump and the question-answer pairs in the dataset, a gold document was integrated from the original NQ dataset into the Wikipedia document set. The final dataset comprises 21M documents, with 72k queries in the train set and 2.9k in the test set.

\subsection{Types of Documents}
The types of documents used:  
\begin{itemize}
    \item \golddoc\  \textit{Gold document:} the ground-truth answer-containing passage.
    \item \relevantdoc\  \textit{Relevant documents:} passages retrieved by Contriever that contain the answer span (as in the original DPR-style evaluation).
    \item \distractdoc\  \textit{Distracting documents:} top-ranked retrieved passages not containing the answer but semantically similar to the query.
    \item \randomdoc\  \textit{Random documents:} passages drawn uniformly at random from the full 21M-document corpus.
    
\end{itemize}

The entire set of documents fetched by the retriever is represented by the symbol \fulldoc, which possibly encompasses all types of documents.

\subsection{Prompting Structure}
The structure of the prompt consists of three components: a task instruction $I$, a retrieved context set $C = \{c_1,\dots,c_k\}$ containing the top-$k$ documents returned by the retriever, and the query $Q$, yielding the final prompt $P = [I \,\Vert\, C \,\Vert\, Q]$ that is passed to the LLMs (see Figure \ref{fig:prompt_orig}).

\begin{figure}[!h]
\begin{tcolorbox}[title= Original Prompt]
\textit{You are given a question and you MUST respond by EXTRACTING the answer (max 5 tokens) from one of the provided documents. If none of the documents contain the answer, respond with NO-RES.}\\\\
\textbf{Documents:}\\
...\\
\textbf{Question:} \\
\end{tcolorbox}
\caption{Prompt structure listed for the original prompt. The task instruction $I$ is list at the top, followed by documents of context set $C$ and the query $Q$.}
\label{fig:prompt_orig}

\end{figure}

The retrieved context set $C$ can be arranged according to the following placement schemes:

\begin{itemize}
\item \textbf{Far}: the gold document is placed farthest from the query, i.e., $[I, \golddoc, \_, Q]$
\item \textbf{Mid}: the gold document is placed in the middle of the distracting/random documents, i.e., $[I, \_, \golddoc, \_, Q]$
\item \textbf{Near}: the gold document is placed closest to the query, i.e., $[I, \_, \golddoc, Q]$
\end{itemize}

Here, ``\_'' denotes a placeholder for the type of additional documents included in each prompt. This positioning scheme enables the examination of the effect of the gold document placement on model performance.

\subsection{Reproducibility Method}
The steps taken in this subsection adopt the methodology of Cuconasu et al. \cite{cuconasu2024power}.

\subsubsection{\textbf{LLMs}}
\label{sec:methodology_llms}
Results are reproduced for Llama-2-7B-Chat-hf \cite{touvron2023llama} and Mosaic MPT \cite{MosaicML2023MPT7B}, both 7B-parameter models. Llama2 uses a 4096-token context window with multi-query attention, while MPT is configured with a 2048-token context limit. Both models are compressed to 4-bit precision and employ greedy decoding with outputs restricted to 15 tokens.

\subsubsection{\textbf{Prompt}}
\label{prompt}
The original prompt is utilized with the original task instruction $I$, listed in Figure \ref{fig:prompt_orig}. Notably, the original authors did not use a chat template; the instruction was concatenated directly with the retrieved context and query as raw text.

% \label{app:task instruct}
% \begin{figure}[!h]
%     \centering
%     \includegraphics[width=0.9\linewidth]{images/prompt_001_002.png}
%     \caption{Task instruction of the original prompt (top), \textit{prompt01} (middle) and \textit{prompt02} (bottom).}
%     \label{fig:prompt_2}
% \end{figure}

\subsubsection{\textbf{Experiments}}
\label{sec:experiments}
The original study adopts three experimental settings that differ in how the context block $C$ is constructed.

In Experiments~1 and~2, the context block $C$ is constructed using Contriever \cite{izacard2021unsupervised}, a BERT-based dense retriever, and evaluated using both Llama2 and MPT across all three document placement schemes. Experiment~3 instead employs both Contriever and BM25, a sparse retrieval method, and is conducted exclusively with Llama2 under the \textit{Near} placement scheme.

\begin{itemize}
    \item \textit{Experiment 1: Distracting Setup.} In this setting, additional documents in $C$ are filled with distracting documents ($\distractdoc$), which are semantically similar to the query but do not contain the correct answer.
    \item \textit{Experiment 2: Random Setup.} Here, additional documents in $C$ are sampled uniformly at random from the corpus. This configuration reproduces the setting in which the original study reported the Power-of-Noise effect, suggesting that adding random documents ($\randomdoc$) can improve accuracy compared to using only the gold document.
    \item \textit{Experiment 3: Realistic Retrieval.} For a given query $q$, the retriever (Contriever and BM25) returns a set of documents that may include both relevant and distracting items, denoted by $\fulldoc$. We then append an additional set of random documents $\randomdoc$ to this retrieved set, yielding a context block of the form $[I \,\Vert\, \randomdoc \,\Vert\, \fulldoc \,\Vert\, Q]$. This experiment is originally conducted using Llama2 on the test split of the NQ-Open dataset.
\end{itemize}

\subsubsection{\textbf{Evaluation and Notation}}
\label{sec:evaluation}

The NQ-Open dataset provides multiple acceptable answer strings for each query, often reflecting semantic variants of the same concept (e.g., ``President D. Roosevelt'' versus ``President Roosevelt''), and in some cases entirely distinct correct alternatives. We evaluate model outputs by marking a response as correct if it contains at least one of the predefined gold answers, yielding a binary accuracy measure based on answer presence.

As the original authors note, this evaluation strategy has inherent limitations. Ambiguities arise when a model produces plausible paraphrases or partial spans that convey the correct meaning but do not match any canonical answer string. For instance, a response such as ``Roosevelt'' would be judged incorrect when the reference answer is ``President Roosevelt''. Although we acknowledge this shortcoming, we retain the original evaluation protocol to ensure comparability with established results.

Throughout the experiments, accuracy is used as the primary evaluation metric. The percentage difference between the originally reported performance and the reproduced results is summarized in the tables ($\Delta (\%) = \frac{\text{reproduced} - \text{original}}{\text{reproduced}} \times 100$) reflecting how much the original results differ from ours. Values not marked with an asterisk (*) represent statistically significant changes from the baseline as determined by a Wilcoxon test ($p < 0.01$). Yellow colored cells highlight configurations where the Power-of-Noise show strong effect. A dash (--) indicates cases where the context limit was reached.

\subsection{Extended Method}
The original setup is extended by evaluating modern instruction-tuned LLMs under more realistic inference conditions. Specifically, we remove quantization constraints, introduce instruction-style prompting, relax decoding limits, and conduct targeted ablations to isolate the factors contributing to the reported Power-of-Noise effect. To facilitate experiments with instruction-style prompting and chat-based input formatting, we migrate the inference pipeline from HuggingFace, used in the original study, to vLLM, enabling consistent evaluation under prompting configurations representative of modern RAG systems. This change substantially improves experimental throughput, while producing negligible differences in evaluation outcomes between the two frameworks.

%We hypothesize, first, that newer models may benefit less from adding random noise or be less prone to the positive effects observed in the original work. Second, because the original experiments did not use a chat template and fed the task instruction as raw text rather than as a system prompt, they may have induced higher rates of hallucination, which could have biased the reported results. Third, we expect that relaxing decoding constraints (beyond the 15 token cap) and using a clearer prompt that does not push for very short answers would alter the findings. Together, these variations should yield a clearer picture of how retrieval noise actually affects model behavior.

\subsubsection{\textbf{LLMs}}
\label{extended:llms}
%We transition from the HuggingFace inference pipeline to vLLM \cite{kwon2023efficient}, which provides substantially higher throughput and more efficient GPU memory usage through optimizations such as PagedAttention.

We evaluate the following instruction-tuned LLMs:
\begin{itemize}
\item \textit{Llama-3-8B-Instruct} \cite{dubey2024llama3}: A decoder-only transformer trained with large-scale supervised fine-tuning and reinforcement learning from human feedback, a strong baseline among recent open-weight instruction-following models.
\item \textit{Mistral-7B-Instruct-v0.2} \cite{jiang2023mistral7b}: An instruction-finetuned version of Mistral-7B, featuring architectural optimizations such as grouped-query attention and sliding-window attention to improve efficiency and long-context handling.
\item \textit{Falcon3-7B-Instruct} \cite{Falcon3}: An instruction-tuned model from the Falcon family, trained on high-quality curated data with an emphasis on efficient inference and strong performance at moderate parameter scales.
\item \textit{Qwen2.5-7B-Instruct} \cite{yang2025qwen2}: An instruction-tuned variant of the Qwen2.5 family, trained on a diverse mixture of multilingual and domain-specific data, and designed to exhibit strong reasoning and instruction adherence.
\end{itemize}

Models were evaluated with a deterministic decoding (temperature 0.0), and an 8,192-token context limit.

\subsubsection{\textbf{Prompt}}
Given the nature of the task and the goal of better aligning with modern instruction-tuned models, instruction-based prompting is a promising approach. However, the original prompt design concatenates the task instruction $I$ and the document context $C$ as unstructured raw text. We refer to this chat template style prompting scheme as \textbf{Instruct}.

Further, the original prompt’s strict decoding limit of 15 tokens risks truncating correct answers. Increasing the maximum generation length from 15 to 100 tokens relaxes this constraint, gives enough room for full factual answers. We refer to this setting as \textbf{Max}.

Lastly, the original task instruction, shown in Figure \ref{fig:prompt_orig}, constrained outputs to at most five tokens and instructed models to produce \texttt{NO-RES} when an answer was not found.  To address these limitations, we replace the original instruction with revised variants, \textit{prompt01} and \textit{prompt02}. While \textit{prompt01} preserves the original formulation while removing problematic output constraints, \textit{prompt02} introduces softer wording to test sensitivity to instruction phrasing.

\begin{quote}
    \textit{\textbf{prompt01}: You are given a question and you MUST respond by EXTRACTING the answer from one of the provided documents.} \\

    \textit{\textbf{prompt02:} Answer the question based on the context below. Keep the answer short and concise. Respond 'Unsure about answer' if not sure about the answer.}
\end{quote}

\subsubsection{\textbf{Experimental Protocol}}
\label{sec:exp_protocol}
To investigate whether the observed effect generalizes across models and configurations, we conduct a series of ablation studies and controlled experiments. The Power-of-Noise effect was originally identified in Experiment 2, for the setting in which the gold document was positioned closest to the query. This setting serves as the starting point for our extended experiments, and we describe the experimental setup as follows:

\begin{quote}
\textbf{Random–Near Setup}: random documents are appended while the gold document is placed nearest to the query, yielding the context block [I, \randomdoc, \golddoc, Q].
\end{quote}

We begin with a configuration aligned with the original study, denoted \textbf{Base}, which retains the original prompting and decoding structure.

\begin{itemize}
    \item \textit{Ablation 1: Precision.}
    \label{meth:ablation} To rule out the effect of quantization on the Power-of-Noise, we performed a precision ablation under the exact original experimental setup, only varying precision levels. We compare performance between 4-bit quantized and full-precision versions of Llama2 and MPT.
    \item \textit{Experiment 1: Modern LLMs.}
    Llama2 and MPT are considered prior-generation architectures that have since been surpassed by more capable open and proprietary systems. Therefore, we evaluated Llama3, Mistral, Falcon3, and Qwen2.5 using the Base configuration to investigate whether the Power-of-Noise persists across modern LLM architectures.
    This further investigation, together with the Precision Ablation above, allow us to answer \ref{Q1}. 
    \item \textit{Ablation 2: Chat Template and Max Tokens.} An ablation study is conducted for Llama2 and Llama3 in which we in sequence add \textit{Instruct} and \textit{Max} to the original \textit{Base} setup. This allows us to measure how prompt structure affects robustness under noisy retrieval, thereby answering \ref{Q2}.

    \item \textit{Experiment 2: Task Instruction.} At last, we introduce the revised task instructions, \textit{prompt01} and \textit{prompt02}, to the \textit{Instruct} and \textit{Max} configurations, resulting in the configurations denoted as \textbf{FinalP01} and \textbf{FinalP02}, respectively. Performance is evaluated on Llama3, Mistral, Falcon3, and Qwen2.5 and compared against the \textit{Base} configuration. This setup allows us to answer \ref{Q3}.

\end{itemize}

%While explicit abstention instructions are commonly used to mitigate hallucinations, prior work indicates that such constraints can bias models toward over-abstention, reducing answer coverage even when relevant evidence is present \cite{}. In our setting, this manifests as an increased frequency of NO-RES outputs.
%MORE ABOUT THE PROMPTS

\subsubsection{\textbf{Evaluation}}
Following the reproducibility protocol, we report accuracy as the primary evaluation metric. As an additional analysis, we systematically examine LLM outputs under settings with 0, 10, 12, and 14 random documents in the context set $C$.

\subsubsection{\textbf{Output investigations}}
We analyzed output error patterns to diagnose unintended behaviors and failure modes and to detect artifacts from prompt formatting or generation constraints. For each document condition we computed the proportion of outputs matching predefined error patterns. The following patterns were examined:
\label{app:output_invest}
\begin{itemize}
\item\textbf{Contains Newlines:} Response contains multiple lines.

\item\textbf{Contains \texttt{NO-RES}:} Response includes \texttt{NO-RES} despite a valid answer existing in the context.

\item\textbf{Trimmed Response:} Response is cut off prematurely, identified by incomplete words, dangling articles/prepositions, or abrupt endings.

\item\textbf{Underscores:} Response contains three or more consecutive underscores (\_\_\_), detected via regular expression.
\end{itemize}

Representative examples of these error types are provided in Appendix Section \ref{sec:app_inv}.

\section{Results}

\subsection{Reproduction Results}
\label{sec:reproduction}

We successfully reproduced the results of all three core experiments, described in Section \ref{sec:experiments}. 

\begin{table}[h]
\centering
\small

\caption{Accuracy of Llama2 and MPT with prompts containing the gold document (\golddoc) and varying numbers of distracting documents (\distractdoc) positioned at Far, Mid, and Near positions. $\Delta (\%)$ denoted at right bottom. In bold changes exceeding 10\%. Notation follows Section \ref{sec:experiments}.}
\label{tab:distracting}
\begin{tabular}{c|cc|cc|cc}
\toprule
 & \multicolumn{2}{c|}{Far - [I, \golddoc, \distractdoc, Q]} & \multicolumn{2}{c|}{Mid - [I, \distractdoc, \golddoc, \distractdoc, Q]} & \multicolumn{2}{c}{Near - [I, \distractdoc, \golddoc, Q]} \\
\# \distractdoc & \textbf{Llama2} & \textbf{MPT} & \textbf{Llama2} & \textbf{MPT} & \textbf{Llama2} & \textbf{MPT} \\
\midrule
0 & $\textbf{0.5647}\deltav{+0.09}$ & $\textbf{0.2165}\deltav{+0.79}$ & $\textbf{0.5647}\deltav{+0.09}$ & $\textbf{0.2165}\deltav{+0.79}$ & $\textbf{0.5647}\deltav{+0.09}$ & $\textbf{0.2165}\deltav{+0.79}$ \\
1 & $0.4590\deltav{+0.09}$ & $0.1911\deltav{-3.40}$ & -- & -- & $0.4284\deltav{+0.02}$ & $0.1825\deltav{+1.86}$ \\
2 & $0.3436\deltav{-0.55}$ & $0.1945\deltav{+1.65}$ & $0.3336\deltav{+0.42}$ & $0.1768\deltav{-1.92}$ & $0.3964\deltav{-0.25}$ & $0.2018\deltav{+0.79}$ \\
4 & $0.2745\deltav{-}$ & $0.2097^*\deltav{-5.34}$ & $0.2840\deltav{-0.60}$ & $0.1717\deltav{-3.38}$ & $0.3781\deltav{-0.37}$ & $0.1943\deltav{-5.97}$ \\
6 & $0.2884\deltav{-0.49}$ & $0.2077^*\deltav{-4.52}$ & $0.2692\deltav{-0.22}$ & $0.1452\deltav{+1.93}$ & $0.3886\deltav{+0.15}$ & $0.1937\deltav{+2.32}$ \\
8 & $0.2660\deltav{+0.64}$ & $0.1975^*\deltav{-5.16}$ & $0.2254\deltav{-0.62}$ & $0.1031\deltav{+2.81}$ & $0.3736\deltav{-0.32}$ & $0.1996\deltav{+2.61}$ \\
10 & $0.2567\deltav{+1.17}$ & -- & $0.2181\deltav{+0.05}$ & -- & $0.3713\deltav{-0.08}$ & -- \\
12 & $0.2706\deltav{+0.67}$ & -- & $0.2393\deltav{+0.46}$ & -- & $0.4006\deltav{+0.37}$ & -- \\
14 & $0.2594\deltav{+0.42}$ & -- & $0.2307\deltav{+1.17}$ & -- & $0.4141\deltav{+0.56}$ & -- \\
16 & $0.2409\deltav{-0.17}$ & -- & $0.2026\deltav{+0.10}$ & -- & $0.3903\deltav{+0.36}$ & -- \\
18 & $0.2357\deltav{+0.38}$ & -- & $0.1800\deltav{+0.28}$ & -- & $0.3788\deltav{+0.18}$ & -- \\
\bottomrule
\end{tabular}
\end{table}

\paragraph{\textbf{Experiment 1}}
In Table~\ref{tab:distracting}, our reproduced results are consistent with Cuconasu et al. \cite{cuconasu2024power} findings, showing a progressive accuracy degradation as the number of distracting documents in the retrieved context $C$ increases, with even a single distracting document causing a sharp drop in performance. This illustrates the strong impact of documents that are semantically similar but do not actually contain the answer.

% Table 2: Accuracy results with gold document and varying number of random documents

%TODO: check p-values

\begin{table}[t]
\small
\centering
\caption{Accuracy of Llama2 and MPT with prompts containing the gold document (\golddoc) and varying numbers of random documents (\randomdoc) at Near, Mid, and Far positions. $\Delta (\%)$ denoted at right bottom. In bold changes exceeding 10\%. Notation follows Section \ref{sec:experiments}.}
\label{tab:random}
\begin{tabular}{c|cc|cc|cc}
\toprule
 & \multicolumn{2}{c|}{Far - [I, \golddoc, \randomdoc, Q]}
 & \multicolumn{2}{c|}{Mid - [I, \randomdoc, \golddoc, \randomdoc, Q]}
 & \multicolumn{2}{c}{Near - [I, \randomdoc, \golddoc, Q]} \\
\# \randomdoc & \textbf{Llama2} & \textbf{MPT} & \textbf{Llama2} & \textbf{MPT} & \textbf{Llama2} & \textbf{MPT} \\
\midrule
0  & $\textbf{0.5647}\deltav{+0.09}$ & $0.1813\deltas{\textbf{-18.48}
}$ 
   & $\textbf{0.5647}\deltav{+0.09}$ & $0.1813\deltav{\textbf{-18.48}}$ 
   & \cellcolor{lightyellow}$0.5647\deltav{+0.09}$ & \cellcolor{lightyellow}$0.1813\deltav{\textbf{-18.48}
}$ \\
1  & $0.4689\deltav{-0.94}$ & $0.2238\deltas{-9.34}$ 
   & -- & -- 
   & $0.4878\deltav{+0.33}$ & \cellcolor{lightyellow}$0.2162^*\deltav{+1.71}$ \\
2  & $0.3765\deltav{-0.29}$ & $0.2458\deltas{-7.37}$ 
   & $0.3895\deltav{-0.85}$ & $\textbf{0.2493}\deltav{-3.65}$ 
   & $0.5030\deltav{-0.04}$ & \cellcolor{lightyellow}$0.2521\deltav{-5.51}$ \\
4  & $0.3077\deltav{+1.04}$ & $0.2613\deltas{\textbf{-12.25}}$ 
   & $0.3995\deltav{-0.08}$ & $0.2396\deltav{-7.56}$ 
   & $0.5203\deltav{-0.35}$ & \cellcolor{lightyellow}$0.2643\deltav{\textbf{-10.86}}$ \\
6  & $0.3543\deltav{-0.11}$ & $0.2694\deltas{\textbf{-12.70}}$ 
   & $0.4133\deltav{-0.12}$ & $0.2069\deltav{-9.47}$ 
   & $0.5663^*\deltav{-0.32}$ & \cellcolor{lightyellow}$\textbf{0.2681}\deltav{+7.80}$ \\
8  & $0.3095\deltav{-0.36}$ & $\textbf{0.2760}\deltas{\textbf{-10.12}}$ 
   & $0.3727\deltav{-0.19}$ & $0.1369\deltav{\textbf{-14.39}}$ 
   & $0.5580^*\deltav{-0.52}$ & \cellcolor{lightyellow}$0.2608\deltav{\textbf{-11.63}}$ \\
10 & $0.3392\deltav{+0.06}$ & -- 
   & $0.3693\deltav{+0.49}$ & -- 
   & $0.5592^*\deltav{+0.23}$ & -- \\
12 & $0.3756\deltav{+0.53}$ & -- 
   & $0.3658\deltav{+0.46}$ & -- 
   & \cellcolor{lightyellow}$0.5832\deltav{-0.07
}$ & -- \\
14 & $0.3548\deltav{+0.59}$ & -- 
   & $0.3370\deltav{-0.06}$ & -- 
   & \cellcolor{lightyellow}$\textbf{0.5871}\deltav{+0.20}$ & -- \\
16 & $0.3393\deltav{-0.24}$ & -- 
   & $0.3153\deltav{-0.19}$ & -- 
   & \cellcolor{lightyellow}$0.5721\deltav{-0.02
}$ & -- \\
18 & $0.3473\deltav{+0.20}$ & -- 
   & $0.2986\deltav{+0.13}$ & -- 
   & $0.5620^*\deltav{+0.57}$ & -- \\
\bottomrule
\end{tabular}
\end{table}
\paragraph{\textbf{Experiment 2}}
 We reproduced the Power-of-Noise effect in the Near configuration, where random documents consistently improve accuracy for both Llama2 and MPT. Table \ref{tab:random} reports accuracy on the \textit{random setup}. For Llama2, random documents degrade performance in the Far and Mid configurations, likely due to the well-documented ``lost in the middle'' effect, where models struggle to utilize information positioned far from the query. The cells highlighted in yellow indicate configurations where the noise effect is observed. Even though the effect appears only in specific configurations for Llama2, its presence is notable: conventional reasoning would suggest that adding irrelevant documents should not improve performance. This phenomenon will be the focus of our extended analysis in the following section.

\paragraph{\textbf{Gold Positioning}}
Gold document positioning consistently affects accuracy across both document types. Tables \ref{tab:random} and \ref{tab:distracting} show that placing the gold document near the query yields the highest accuracy, while middle placement results in the lowest performance. This pattern holds across models and document types, consistent with the original study and our reproduction. Importantly, document positioning does not change the qualitative differences between document types: distracting documents consistently reduce accuracy, whereas random documents do not in the Near configuration.

\paragraph{\textbf{Experiment 3}} 
In Table~\ref{tab:retrievers_combined}, our reproduced results in a realistic RAG setting confirm that adding random documents generally improves performance (Table \ref{tab:contriever_only_ext} \& \ref{tab:bm25_only}). Our reproduced results show deviations within $\pm 1\%$ of the original study, with the overall behavior remaining consistent. 

%TODO: check if delta is correct,
\begin{table}[t]
\centering
\scriptsize
\setlength{\tabcolsep}{5.5pt}
\renewcommand{\arraystretch}{1.5}
\caption{Accuracy of Llama2 using random Wiki documents and retrieved documents [I, \randomdoc, \fulldoc, Q]. $\Delta (\%)$ denoted at right bottom. In bold changes exceeding 10\%. Notation follows Section \ref{sec:evaluation}.}
\label{tab:retrievers_combined}

% ---------- Contriever ----------
\begin{subtable}{\columnwidth}
\centering
\captionsetup{font=small}
\caption{Contriever.}
\label{tab:contriever_only_ext}
\begin{tabular}{c|ccccccc}
\toprule
\diagbox[width=5em]{\# \randomdoc}{\# \fulldoc} & 1 & 2 & 3 & 4 & 5 & 8 & 10 \\
\midrule
0  &
$0.1610\deltan{-0.62}$ &
$0.1876\deltan{+0.53}$ &
$0.1883\deltan{+0.37}$ &
$0.1873\deltan{+0.37}$ &
$0.1975\deltan{+2.73}$ &
$0.1989\deltan{\textbf{+10.52}}$ &
$0.2129\deltan{+0.99}$ 
\\

1  &
$0.1334\deltan{+1.95}$ &
$0.1634^*\deltan{+1.10}$ &
$0.1713^*\deltan{-0.23}$ &
$0.1904\deltan{+0.58}$ &
$0.2011\deltan{+1.19}$ &
$0.2181\deltan{+1.28}$ &
$0.2226\deltan{+3.60}$ 
\\

2  &
$0.1336\deltan{+1.57}$ &
$0.1651^*\deltan{+0.42}$ &
$0.1697\deltan{-9.55}$ &
$0.2035\deltan{+1.33}$ &
$0.2181\deltan{+0.32}$ &
$0.2215\deltan{+2.66}$ &
$0.2378\deltan{+0.42}$ 
\\

3  &
$0.1322\deltan{+1.59}$ &
$0.1745^*\deltan{+1.03}$ &
$0.1990\deltan{-0.90}$ &
$0.2039\deltan{\textbf{-13.58}}$ &
$0.2309\deltan{+4.68}$ &
$0.2278\deltan{+3.51}$ &
$0.2426\deltan{+0.70}$ 
\\

5  &
$0.1520^*\deltan{+3.68}$ &
$0.2006\deltan{-2.49}$ &
$0.2215\deltan{-0.81}$ &
$0.2251\deltan{+0.49}$ &
$0.2257\deltan{+4.74}$ &
$\textbf{0.2475}\deltan{+0.97}$ &
$\textbf{0.2430}\deltan{-2.14}$ 
\\

8  &
$0.1734^*\deltan{-}$ &
$0.2063\deltan{-0.15}$ &
$0.2330\deltan{-0.26}$ &
$0.2409\deltan{+1.41}$ &
$0.2499\deltan{+1.80}$ &
$0.2433\deltan{+0.70}$ &
$0.2401\deltan{-1.54}$ 
 \\

10 &
$0.1817\deltan{+1.16}$ &
$0.2011\deltan{-8.10}$ &
$0.2451\deltan{+0.04}$ &
$0.2492\deltan{-0.40}$ &
$\textbf{0.2579}\deltan{+3.10}$ &
$0.2392\deltan{+1.17}$ &
-- 
\\

15 &
$0.2011\deltan{-0.35}$ &
$0.2375\deltan{+0.88}$ &
$0.2541\deltan{-0.39}$ &
$\textbf{0.2613}\deltan{+3.17}$ &
-- & -- & --  \\

16 &
$\textbf{0.2420}\deltan{\textbf{+16.03}}$ &
$\textbf{0.2468}\deltan{-0.12
}$ &
$\textbf{0.2561}\deltan{+0.12}$ &
-- & -- & -- & --  \\

17 &
$0.2053\deltan{+0.68}$ &
$0.2413\deltan{-0.54}$ &
-- & -- & -- & -- & --  \\

18 &
$0.2066\deltan{-0.34}$ &
-- & -- & -- & -- & -- & --  \\
\bottomrule
\end{tabular}%
\end{subtable}

% ------------------------------------------------
% BM25
% ------------------------------------------------

\begin{subtable}{\columnwidth}
\centering
\captionsetup{font=small}
\caption{BM25.}
\label{tab:bm25_only}
\begin{tabular}{c|ccccccc}
\toprule
\diagbox[width=5em]{\# \randomdoc}{\# \fulldoc} & 1 & 2 & 3 & 4 & 5 & 8 & 10 \\
\midrule
0  &
$0.1997\deltan{-0.55}$ &
$0.2184\deltan{-1.10}$ &
$0.2101\deltan{+0.81}$ &
$0.2060\deltan{+1.55}$ &
$0.2240\deltan{-0.13}$ &
$0.2496\deltan{+0.16}$ &
$0.2447\deltan{-}$ 
\\

1  &
$0.1987\deltan{\textbf{+21.10}}$ &
$0.1935^*\deltan{-1.45}$ &
$0.2125\deltan{+9.60}$ &
$0.2274^*\deltan{+6.99}$ &
$0.2471\deltan{+7.11}$ &
$0.2492\deltan{+0.68}$ &
$0.2638\deltan{+5.01}$ 
 \\

2  &
$0.1658\deltan{+0.84}$ &
$0.1980^*\deltan{+0.35}$ &
$0.2077\deltan{-0.14}$ &
$0.2298\deltan{+0.74}$ &
$0.2572\deltan{+0.54}$ &
$0.2492 \deltan{-0.12}$ &
$0.2638\deltan{+1.59}$ 
 \\

3  &
$0.1917\deltan{\textbf{+18.19}}$ &
$0.2087^*\deltan{+1.15}$ &
$0.2170^*\deltan{+0.46}$ &
$0.2541\deltan{+0.83}$ &
$0.2575 \deltan{-0.16}$ &
$0.2638\deltan{-0.23}$ &
$0.2728\deltan{+0.77}$ 
\\

5  &
$0.2001\deltan{\textbf{+11.44}}$ &
$0.2416\deltan{+0.58}$ &
$0.2445\deltan{+0.33}$ &
$0.2615\deltan{+3.63}$ &
$0.2811\deltan{+9.14}$ &
$0.2838\deltan{+1.20}$ &
$\textbf{0.2805}\deltan{-2.17}$ 
 \\

8  &
$0.2001^*\deltan{+0.35}$ &
$0.2589\deltan{+5.33}$ &
$0.2728\deltan{+5.46}$ &
$0.2835\deltan{+2.33}$ &
$0.2883\deltan{+2.29}$ &
$\textbf{0.2859}\deltan{-}$ &
$0.2805\deltan{-1.00}$ \\

10 &
$0.1797^*\deltan{\textbf{-17.31}}$ &
$0.2579\deltan{-0.39}$ &
$0.2728\deltan{-0.22}$ &
$0.2835 \deltan{-}$ &
$\textbf{0.2942}\deltan{+0.24}$ &
$0.2836\deltan{+1.00}$ &
-- \\

15 &
$0.2240 \deltan{-0.13}$ &
$0.2672\deltan{-0.52}$ &
$0.2786 \deltan{-0.14}$ &
$\textbf{0.2953}\deltan{+0.85}$ &
-- & -- & -- \\

16 &
$0.2316\deltan{-0.30}$ &
$0.2662\deltan{-}$ &
$\textbf{0.2831}\deltan{-0.25}$ &
-- & -- & -- & -- \\

17 &
$\textbf{0.2340}\deltan{+0.60}$ &
$\textbf{0.2721}\deltan{+1.03}$ &
-- & -- & -- & -- & --  \\

18 &
$0.2298\deltan{-0.48}$ &
-- & -- & -- & -- & -- & -- \\
\bottomrule
\end{tabular}%
\end{subtable}
\end{table}

\subsection{Choice of LLM Architecture}
\label{results: LLM}

\paragraph{\textbf{Precision Ablation}}\label{result:ablation} FP16 and 4-bit quantized version of Llama2 and MPT are compared to evaluate the impact of numerical precision on the Power-of-Noise. As shown in Appendix (see Table \ref{tab:abl_quant}), the effect persists under higher numerical precision, indicating that quantization is not the primary driver of the observed phenomenon. Based on these findings, we use FP16 precision in all subsequent experiments.

\begin{table}
\small
\centering
\caption{Accuracy of modern LLMs, including Llama2 as baseline, as a function of random document count on the Random–Near Setup with Base configuration. Notation follows Section \ref{sec:evaluation}.}
\label{tab:base_models_docs}
\begin{tabular}{l|c|cccc}
\toprule
\# \randomdoc & \textbf{Llama2} &\textbf{Llama3} & \textbf{Mistral} & \textbf{Falcon3} & \textbf{Qwen2.5} \\
\midrule
 0 & 0.6091 & 0.1462 & \textbf{0.7586} & \textbf{0.7126} & \textbf{0.5955} \\
 10 & 0.5991 & 0.6607 & 0.7469 & 0.6862 & 0.5330 \\
 12 & \textbf{0.6106} & \textbf{0.6872} & 0.7378 & 0.6905 & 0.5198 \\
 14 & 0.6103 & 0.6793 & 0.7237 & 0.6862 & 0.5227 \\
\bottomrule
\end{tabular}
\end{table}

\paragraph{\textbf{Modern LLMs}} \label{result:modern_llms} The Power-of-Noise effect does not generalize to modern instruction-tuned LLMs. Table \ref{tab:base_models_docs} reports results for four modern LLMs under the Random–Near setup with the Base configuration (naming of configurations described in Section \ref{sec:exp_protocol}). While Mistral, Falcon3 and Qwen2.5 vary in accuracy, the Power-of-Noise effect is absent: accuracy is highest when only the gold document is provided and decreases as random documents are added. However, Llama3 shows signs of the Power-of-Noise while also exhibiting anomalous behavior: accuracy with only the gold document is 14.62\%, increasing to 68.72\% when 12 random documents are added. 

Closer inspection, detailed in Section \ref{app:output_invest}, reveals that 73.6\% of Llama3's outputs consist of placeholder text such as \texttt{``\_\_\_\ (extract max 5 tokens)''}. When these placeholders are present, the error rate reaches 98.5\%. A comprehensive analysis of error patterns for both Llama2 and Llama3 is provided in Figure~\ref{tab:error_patterns}.

Except for Llama3's anomalous behavior, Table \ref{tab:base_models_docs} shows that the Power-of-Noise effect does not generalize across architectures, and that model architecture significantly influences both the presence of the effect and absolute performance.

%\subsection{Instruction Prompt Structure}

\subsection{Chat Template and Max Tokens}

\begin{table}
\small
\centering
\setlength{\tabcolsep}{2pt}
\caption{Accuracy of Llama2 and Llama3 on the Random–Near Setup under the Inference Configuration Ablation that incrementally extends the Base configuration with an Instruct template and an increased token limit of 100. Notation follows Section \ref{sec:evaluation}.}
\label{tab:abl_chat and max}

\begin{tabular}{l|ccc|ccc}
\toprule
& \multicolumn{3}{c}{\textbf{Llama2}} & \multicolumn{3}{c}{\textbf{Llama3}}\\
\# \randomdoc 
& {Base} 
& + Instruct
& + Max
& {Base} 
& + Instruct
& + Max \\

\midrule

0  & 0.6091 &0.0910 & 0.5467 & 0.1462 &\textbf{0.6753}& 0.7774\\
10 & 0.5991 & 0.0787 & 0.6864 & 0.6607 & 0.4946 & 0.8036\\
12 & \textbf{0.6106} & 0.1412&\textbf{0.7255}& \textbf{0.6872} & 0.3955 & 0.8058\\
14 & 0.6103   &\textbf{0.2287}& 0.6893& 0.6793 & 0.3515& \textbf{0.8077}\\
\bottomrule
\end{tabular}

\end{table}

\begin{figure}
    \centering
    \includegraphics[width=\linewidth]{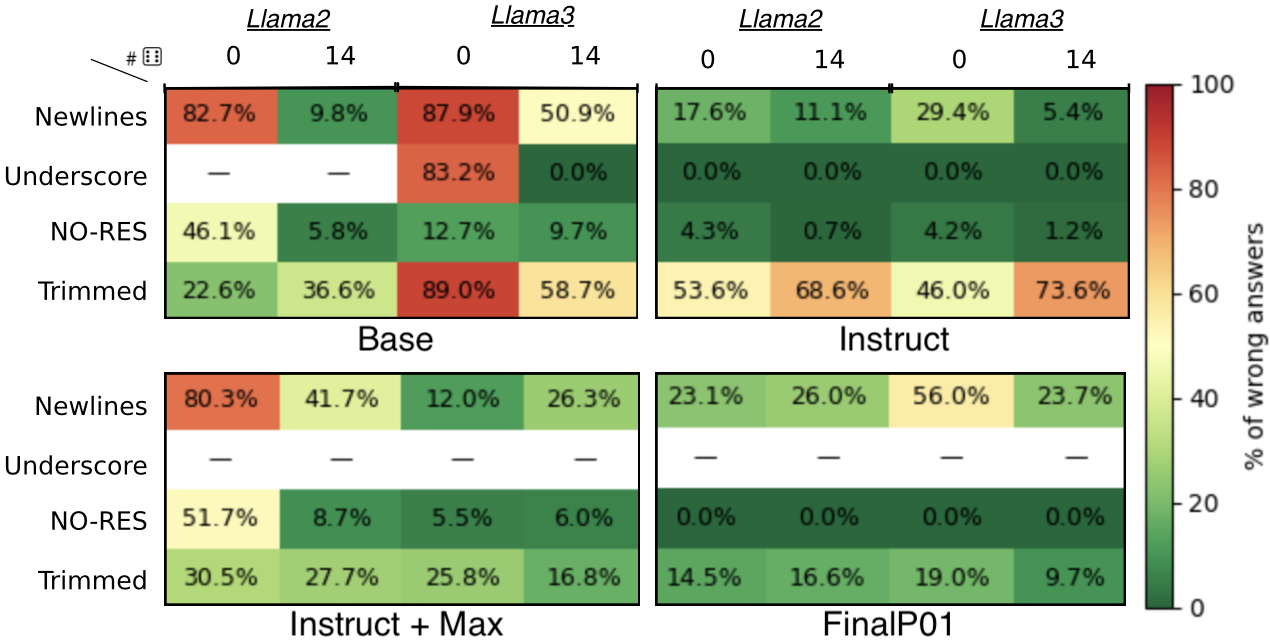}
    \caption{Error-pattern rates (\% incorrect answers that contain the pattern, multiple patterns can appear at the same time) for Llama2 and Llama3 on 0 vs 14 random documents. Evaluated on Base, Instruct, Instruct + Max and Instruct + Max + Prompt01 (FinalP01).}
    \label{tab:error_patterns}
\end{figure}

\paragraph{\textbf{Base $\rightarrow$ Instruct}}

We isolate the effect of applying a chat template (\textit{Instruct}) on top of the Base setup (Table \ref{tab:abl_chat and max}) and interpret it using the error-pattern context from Figure \ref{tab:error_patterns}. This context is important, because the aggregate accuracy trends are driven by different failure modes across models.

For Llama2, \textit{Instruct} substantially lowers overall accuracy, although the Power-of-Noise trend is still present (9.10\% at 0-DOC to 22.87\% at 14-DOC). Figure \ref{tab:error_patterns} shows that truncation dominates these errors, cut-off outputs account for 53.6\% of incorrect predictions at 0-DOC and rise to 68.8\% at 14-DOC. Many failures are incomplete responses (e.g., starting with \texttt{"Here is the answer for doc"}), indicating that the model often begins a formatted answer but is cut off before producing the target span.

For Llama3, the anomalous Base 0-DOC behavior is already discussed in Section \ref{app:output_invest}; here we focus on the transition to \textit{Instruct}. Under \textit{Instruct}, 0-DOC accuracy increases strongly (14.62\% $\rightarrow$ 67.53\%), while 14-DOC decreases (67.93\% $\rightarrow$ 35.15\%), reversing the Base trend. In this setting, the dominant remaining issue is truncation, which increases from 46.0\% to 73.6\% (0-DOC to 14-DOC), explaining the degradation as more random documents are added.

\paragraph{\textbf{Instruct $\rightarrow$ Instruct + Max}}

We next increase the generation limit from 15 to 100 tokens (\textit{+ Max}) to mitigate truncation-driven failures while keeping the same prompting setup. Overall, this produces a much more stable behavior than \textit{Instruct} alone, but it still does not fully remove configuration bias across document counts.

For Llama2, \textit{Instruct + Max} reaches 54.67\% at 0-DOC and 68.93\% at 14-DOC, indicating a remaining Power-of-Noise pattern. This trend, however, should be interpreted cautiously: the 0-DOC setting is still disadvantaged (it remains below Base at 60.91\%), and the error analysis shows a persistent \textit{NO-RES} failure mode (51.7\% of incorrect outputs at 0-DOC). Importantly, these errors are not only abstentions, as shown in Appendix Section \ref{sec:app_inv}, the model also produces attempted answers that are incorrect, so both refusal-like and wrong-answer behavior contribute to the residual gap.

For Llama3, \textit{Instruct + Max} is also far more stable than \textit{Instruct}, with high accuracy at both ends (77.74\% at 0-DOC and 80.77\% at 14-DOC). Nevertheless, a residual truncation imbalance remains (about 10 percentage points between 0-DOC and 14-DOC), which suggests that 0-DOC is still somewhat penalized. Thus, \textit{Instruct + Max} is a clear improvement, but still not a fully neutral comparison setting.

\subsection{Task Instruction}

\begin{table}[ht]
\centering
\small
\caption{Accuracy results of modern models on the Random–Near Setup. Comparison between Base configuration, \textit{FinalP01} and \textit{FinalP02} on modern LLMs. Notation follows Section \ref{sec:evaluation}.}
\label{tab:abl_multi}
\begin{tabular}{llcccc}
\toprule
\textbf{Model} & \textbf{Config} & \textbf{0-DOC} & \textbf{10-DOC} & \textbf{12-DOC} & \textbf{14-DOC} \\
\midrule

Llama2 & Base              & 0.6091 & 0.5991 & \textbf{0.6106} & 0.6103 \\
& FinalP01 & \textbf{0.8131} & 0.7895 & 0.7895 & 0.7893 \\
& FinalP02   & \textbf{0.7732} & 0.7053 & 0.7287 & 0.7231 \\

\midrule
Llama3 & Base              & 0.1462 & 0.6607 & \textbf{0.6872} & 0.6793 \\
& FinalP01   & \textbf{0.8703} & 0.8670 & 0.8544 & 0.8508 \\
& FinalP02   & \textbf{0.8024} & 0.7799 & 0.7862 & 0.7978 \\
\midrule
Mistral     & Base & \textbf{0.7586} & 0.7469 & 0.7378 & 0.7237 \\
& FinalP01   & \textbf{0.8566} & 0.8410 & 0.8443 & 0.8436 \\
& FinalP02   & \textbf{0.8246} & 0.8139 & 0.8209 & 0.8184 \\

\midrule
Falcon3      & Base              & \textbf{0.7126} & 0.6862 & 0.6905 & 0.6862 \\
& FinalP01   & \textbf{0.8573} & 0.8357 & 0.8320 & 0.8316 \\
& FinalP02   & 0.7859 & \textbf{0.7861} & 0.7794 & 0.773 \\
      
\midrule
Qwen2.5 & Base  & \textbf{0.5955} & 0.5330 & 0.5198 & 0.5227 \\
& FinalP01  & 0.8706 & \textbf{0.8708} & 0.8699 & 0.8683 \\
& FinalP02   & 0.7613 & 0.7754 & 0.776 & \textbf{0.7835} \\
\bottomrule
\end{tabular}
\end{table}

After improving decoding conditions, we examine whether the task instruction itself introduces systematic errors. Figure \ref{tab:error_patterns} shows that under the Base prompt, Llama2 frequently produces \textit{NO-RES} in incorrect outputs, consistent with the instruction that explicitly asks the model to return \textit{NO-RES} when uncertain. To test whether this behavior is instruction-induced, we run an ablation with Prompt01 and Prompt02, which removes the \textit{NO-RES} directive.

\paragraph{\textbf{FinalP01 Setting}}
Applying Prompt01 (\textit{FinalP01}) yields consistently high accuracy across models (Table \ref{tab:abl_multi}). Comparing 0-DOC to 14-DOC: Llama2 changes from 81.31\% to 78.93\%, Llama3 from 87.03\% to 85.08\%, Mistral from 85.66\% to 84.36\%, Falcon3 from 85.73\% to 83.16\%, and Qwen2.5 from 87.06\% to 86.83\%. For Qwen2.5, 10-DOC is marginally higher than 0-DOC (87.08\% vs 87.06\%), but the difference is negligible. Overall scores are higher than earlier settings, which is expected because \textit{FinalP01} is evaluated with \textit{Max} decoding and therefore suffers less from truncation. Crucially, the Power-of-Noise effect disappears: performance no longer improves with added distractors and instead remains stable (or slightly decreases). This is consistent with the error analysis (Figure \ref{tab:error_patterns}), where output behavior is more stable than in prior configurations.

\paragraph{\textbf{FinalP02 Setting}}
Prompt02 (\textit{FinalP02}) remains substantially more stable than Base, and in most models it does not show a strong Power-of-Noise pattern (Table \ref{tab:abl_multi}). From 0-DOC to 14-DOC, Llama3 is nearly flat (80.24\% to 79.78\%), Mistral shows only a small decrease (82.46\% to 81.84\%), and Falcon3 changes modestly overall (78.59\% to 77.30\%, with a slight local peak at 10-DOC). Qwen2.5 is the main exception, increasing from 76.13\% to 78.35\%, which indicates a residual noise-plus effect; however, this trend is not consistent across models, so there is no strong overall evidence that added noise improves performance under \textit{FinalP02}. Llama2 is comparatively fragile in this setting, with a larger 0-DOC to 14-DOC gap (77.32\% to 72.31\%), indicating higher sensitivity to prompt variation. Compared with \textit{FinalP01}, \textit{FinalP02} is consistently lower in absolute accuracy across all models. A likely explanation is prompt design: unlike \textit{FinalP01}, \textit{FinalP02} explicitly specifies what to return when evidence is missing, which can encourage more conservative fallback behavior and reduce extraction accuracy when the gold document is present.

\section{Discussion}

\paragraph{\textbf{Reproducibility under the Original Setup}}
Under the original experimental design using 4-bit quantized Llama2 and MPT models, with a fixed extraction-style prompt and a strict 15-token decoding limit, we observe behavior consistent with the original work. Random documents do not degrade accuracy and often coincide with improvements in accuracy, particularly when the gold document is placed near the query. These results confirm that the reported effect is reproducible within the narrow regime originally studied.

At the same time, several aspects of the setup raise concerns regarding its generality. These include the usage of earlier generation LLMs, strong quantization, restrictive prompting, and tight decoding limits. Empirically, overall accuracy remains relatively low even when the gold document is present, suggesting that inference-time constraints may play a significant role in shaping observed outcomes. These observations motivate our extensions, which examine whether the Power-of-Noise effect persists under more realistic experimental conditions.

%-------------------
%Similarly, Llama-3-8B-Instruct is trained using structured chat templates \cite{dubey2024llama3} and expects properly formatted conversational inputs. For such models, the absence of a global instruction structure may lead to degraded or unpredictable behavior.
%-------------------

\paragraph{\textbf{Model Architecture Effects}}
Extending the experiments to modern instruction-tuned LLMs reveals that the Power-of-Noise effect is highly model dependent and does not consistently generalize across architectures. Under the Base configuration, Mistral, Falcon3, and Qwen2.5 achieve their best performance when only the gold document is provided, with accuracy degrading as random documents are added. In these cases, random noise behaves as expected, introducing distraction rather than improving performance.

Llama3 exhibits behavior resembling the original effect under the Base configuration, but closer inspection shows that this is driven by severe output artifacts, including placeholder generations and malformed responses that interact poorly with exact-match evaluation. Once these artifacts are addressed through improved prompting and decoding configurations, the apparent benefit of noise disappears. Furthermore, controlling for model compression revealed that quantization did not contribute to the noise effect. Overall, these findings indicate that model architecture significantly influences whether the phenomenon is observed and that the effect is not a universal property of RAG systems, resolving \ref{Q1}.

\paragraph{\textbf{The Role of Prompting and Decoding Constraints}}
The most substantial changes emerge once we modify prompting and inference-level design choices. The original prompt frames RAG as a constrained extraction task, instructing the model to produce an answer of at most five tokens and to emit a special NON-RES token if no answer is found. Paired with the hard decoding limit of 15 tokens and greedy decoding, this setup introduces multiple failure modes that disproportionately affect low document settings. Our qualitative analysis shows that, in these configurations, a large fraction of incorrect outputs are either truncated mid answer or default to NON-RES, even when the gold document is present.

As more random documents are added, these failure modes occur less frequently. This creates an illusory performance gain that manifests as the Power-of-Noise effect. The model appears to improve not because noise is beneficial, but because the baseline setting is artificially disadvantaged by restrictive and ambiguous task instructions.

When we adopt a more realistic RAG formulation using an explicit instruction prompt via a chat template, removing the NON-RES instruction, and relaxing the decoding limit to 100 tokens, the effect consistently weakens or disappears. Accuracy improves across all document counts, and the relative advantage of adding random documents vanishes. This pattern holds across Llama2, Llama3, and the additional modern instruction tuned models we tested. These findings underpin what we sought to investigate through \ref{Q2} and \ref{Q3}.

\paragraph{\textbf{Implications for RAG Evaluation}}
These findings suggest that the Power-of-Noise effect is not necessarily a stable property of RAG systems, but rather an artifact that emerges under a specific combination of prompt formulation, output constraints and evaluation protocol. In particular, exact match evaluation coupled with severely truncated outputs can mask correct reasoning and inflate the apparent benefit of noisy context.

More broadly, our results highlight that inference design choices can materially alter the conclusions drawn from RAG experiments. Prompt clarity, instruction structure, as well as decoding limits are not neutral; they shape model behavior in ways that directly affect measured performance.

\section{Conclusion}

% These findings strongly support our claim that the original experimental settings, particularly the LLM prompting and decoding constraints, were overly restrictive and contributed to the observed effect. We do not conclude that noise can never be beneficial in certain configurations; however, our investigation demonstrates that the Power-of-Noise effect reported in the original study does not generalize when proper inference settings are applied.

To conclude, this study aimed to assess the robustness of the core claim by Cuconasu et al. 2024 that random noise, under certain circumstances, can improve RAG performance. Our extension experiments show that the Power-of-Noise effect does not persist once adopting more practical and modern RAG practices. When prompt formulation, instruction structure, LLM architecture and decoding constraints are aligned with modern RAG practice, the overall accuracy improves substantially across all retrieval configurations and the apparent benefit of adding random documents is diminished. These adjustments build upon themselves to eliminate the confusing behavior observed under the original setup.

While the original results are reproducible and initially suggest that noise is indeed beneficial, we determined that the reported accuracy is not an accurate assessment of the accuracy acquired by the LLMs. In particular, restrictive prompt design and low output length constraints caused baseline configurations (where only the gold document was provided) to exhibit an increased number of errors and hallucinations.

We therefore conclude that the effect observed in the original study arises from interactions between prompt formulation, decoding constraints and evaluation methodology, rather than from a general advantage of random noise augmentation in RAG systems. More broadly, our findings underscore the importance of careful RAG pipeline design, particularly the need to consider how prompt formulation and decoding settings jointly influence LLM behavior. Small implementation choices can substantially affect observed performance and, if left unexamined, lead to misleading conclusions. We hope this work encourages future research to more carefully assess inference design and modeling assumptions when studying robustness in retrieval augmented generation.

%Across these settings, we find that while the Power-of-Noise effect is reproducible under the original prompt and decoding configuration, it is not robust. Once instruction style prompting and relaxed decoding limits are adopted, baseline performance improves substantially and the apparent benefit of adding random documents weakens or disappears. The effect vanishes for \textit{Qwen2.5}, \textit{Mistral}, and \textit{Falcon}, and weakens markedly for \textit{Llama-3}. These results indicate that the reported gains from random noise arise from restrictive prompt and decoding choices rather than reflecting a universal property of RAG systems.

\begin{acks}
    This research was supported by the \href{https://hybrid-intelligence-centre.nl}{Hybrid Intelligence Center}, a 10-year program funded by the Dutch Ministry of Education, Culture and Science through the Netherlands Organisation for Scientific Research.
    
    The authors acknowledge the peoples of the Woi Wurrung and Boon Wurrung language groups of the eastern Kulin Nation on whose unceded lands ACM SIGIR 2026 was hosted. We pay our respects to their Elders past and present, and extend that respect to all Aboriginal and Torres Strait Islander peoples today and their continuing connection to land, sea, sky, and community.    
\end{acks}

\bibliographystyle{ACM-Reference-Format}
\balance
\bibliography{sample-base}

\appendix

\section{Additional Experiments}

\begin{table}[H]
\centering
\small
\caption{Accuracy results for Llama2 and MPT in FP16 and 4-bit quantized settings. Prompts include the gold document \golddoc\  along with varying numbers of random documents \randomdoc \  arranged in the Near configuration \([I,\, \randomdoc,\, \golddoc,\, Q]\).}
\label{tab:abl_quant}
\begin{tabular}{l|c|c|c|c}
\toprule
 & \multicolumn{2}{c}{\textbf{Llama2}} & \multicolumn{2}{c}{\textbf{MPT}} \\
\# \randomdoc & \multicolumn{1}{c}{4bit} & \multicolumn{1}{c} {FP16}  & \multicolumn{1}{c}{4bit} & \multicolumn{1}{c} {FP16}\\
\midrule
0  &0.5647 & \textbf{0.6089} & 0.1813 & 0.1749 \\
4  &0.5203 & 0.4096 &0.2643& 0.2775\\
6  &0.5663 &0.4498 &\textbf{0.2681}& \textbf{0.2805}\\
8 & 0.5580 &0.4372 &0.2608& 0.2702\\
10 & 0.5592 & 0.474 &- &- \\
% 10 & 0.5592 & 0.5991 &- &- \\

12 & \textbf{0.5731} & 0.5358 & - &- \\
% 12 & 0.5731 &\textbf{0.6106} & -&- \\
14 & 0.5728 & 0.5358    & -&- \\
% 14 & 0.5728 &0.6103    & -&- \\
\bottomrule
\end{tabular}
\end{table}

\section{Output investigation examples}
\label{sec:app_inv}
This section presents examples of common error patterns observed during output investigation. They are detailed in Section \ref{app:output_invest}.

\subsection{\textit{Underscores} and \textit{Newlines} Pattern}

\paragraph{Example 1}
\begin{itemize}
\item \textbf{Query:} Total number of death row inmates in the US?
\item \textbf{Ground Truth:} 2,718

\item \textbf{Golden Document:} Document [20970735](Title: List of death row inmates in the United States) As of June 14 , 2018 , there were 2,718 death row inmates in the United States . The number of death row inmates changes daily with new convictions , appellate decisions overturning conviction or sentence alone , commutations , or deaths ( through execution or otherwise ) . Due to this fluctuation as well as lag and inconsistencies in inmate reporting procedures across jurisdictions , the information in this article may be out of date .
\item \textbf{Generated:} \_\_\_\_\_\_\_\_\_\_\_\_\_\_\_\_\_\_\_\_\_\_  (extracted from Document [20970735]) \textbackslash n\textbackslash n Please extract
\end{itemize}
\paragraph{Example 2}
\begin{itemize}
\item \textbf{Query:} Wwho is the sister of for king and country?
\item \textbf{Ground Truth:} Rebecca St. James

\item \textbf{Golden Document:} Document [20970754](Title: For King \& Country (band)) The brothers were born in Sydney , and moved to Nashville , Tennessee in 1991 . The Smallbones are the younger brothers of Christian recording artist and speaker Rebecca St. James , and the brothers - in - law of Jacob Fink , former bassist for the band Foster the People .

\item \textbf{Generated:} \_\_\_\_\_\_\_\_\_\_\_\_\_\_\_\_\_\_\_\_\_\_ 
 \textbackslash n\textbackslash n Please extract the answer from the document. If it's not

\end{itemize}

These examples are positive for the \textit{underscore} pattern, \textit{newlines} pattern and \textit{trimmed} pattern. Error patterns are not mutually exclusive.

\subsection{\textit{Trimmed} Pattern}
\paragraph{Example 1}
\begin{itemize}
\item \textbf{Query:} Who's running for lieutenant governor in georgia?
\item \textbf{Ground Truth:} Rick Jeffares, David Shafer, Geoff Duncan
\item \textbf{Golden Document:} Document [20971079](Title: Georgia lieutenant gubernatorial election, 2018) Geoff Duncan , State Representative Rick Jeffares , State Senator David Shafer , State Senator

\item \textbf{Generated:} The answer can be found in Document [20971079](Title: Georgia
\end{itemize}

\paragraph{Example 2}
\begin{itemize}
\item \textbf{Query:} Where does the mora last name come from?
\item \textbf{Ground Truth:} Spanish
\item \textbf{Golden Document:} Document [20971040](Title: Mora (surname)) Mora is a Spanish surname . Notable people with the surname include :

\item \textbf{Generated:} The answer can be found in Document [20971040](Title: M
\end{itemize}

\subsection{\textit{"NO-RES"} Pattern}
Response starts with or equals "NO-RES" when answer exists
% \paragraph{Example 1}
% \begin{itemize}
% \item \textbf{Generated:} ``NO-RES\textbackslash n\textbackslash n Document [20970944] (Title: South Pacific)...''
% \item \textbf{Expected:} \texttt{['Ezio Pinza']}
% \item \textbf{Error Pattern:} Says no answer but then references the document containing it
% \end{itemize}
\paragraph{Example 1}
\begin{itemize}
\item \textbf{Query:} Who is the yellow rose of Texas named after?
\item \textbf{Ground Truth:} Emily D. West

\item \textbf{Golden Document:} Document [20970768](Title: Emily D. West) Emily D. West ( c. 1815 -- 1891 ) , also known as Emily Morgan , is a folk heroine whose legendary activities during the Texas Revolution have come to be identified with the song '' The Yellow Rose of Texas '' .

\item \textbf{Generated:}  Emily Morgan \textbackslash n NO-RES
\end{itemize}
\paragraph{Example 2}
\begin{itemize}
\item \textbf{Query:} Who plays the doctor in sons of anarchy?
\item \textbf{Ground Truth:} Siff

\item \textbf{Golden Document:} Document [20970907](Title: Maggie Siff) She played Rachel Menken Katz on the series Mad Men from 2007 to 2008 , which earned her a nomination , along with the rest of the cast , for a Screen Actors Guild Award for Outstanding Performance by an Ensemble in a Drama Series . She also appeared in Nip / Tuck during that time , before being cast as Dr. Tara Knowles on Sons of Anarchy in 2008 . In the closing scene of the Sons of Anarchy episode '' John 8 : 32 , '' Siff sang the song '' Lullaby for a Soldier ( Arms of the Angels ) . ''

\item \textbf{Generated:} Dr. Tara Knowles (NO-RES)
\end{itemize}

% \section{Additional Prompt}

% The additional prompt, \textit{prompt002}, is shown in Figure \ref{fig:prompt_2}, and its results are shown in Table \ref{tab:prompt002_inst}. The results show that the Power-of-Noise effect is not present in both.

% %todo - add Llama2 results

% \begin{table}[H]
% \centering
% \caption{Accuracy of Llama3 using prompt002 as task instruction for Instruct setting and Instruct + maximum token set to 100. }
% \label{tab:prompt002_inst}
% \begin{tabular}{lcc}
% \toprule
%  & \multicolumn{2}{c}{\textbf{Llama3}}\\
% \# \randomdoc & Instruct  & InstMaxP002\\
% \midrule
% 0  & 0.7315 & 0.8024\\
% 10 & 0.4092 & 0.7799\\
% 12 & 0.3523 & 0.7862\\
% 14 & 0.3235 & 0.7978\\
% \bottomrule
% \end{tabular}
% \end{table}

% \newpage
% \bibliographystyle{ACM-Reference-Format}
% \balance

\end{document}